\documentclass[epj]{svjour}
\usepackage{graphics}
\begin{document}
\title{Knockout of proton-neutron pairs from $^{16}$O with electromagnetic probes}
\author{D.G. Middleton\inst{1} \and J.R.M. Annand\inst{2} \and C. Barbieri\inst{3}  \and C. Giusti\inst{4} \and P. Grabmayr\inst{1} \and T. Hehl\inst{1} \and I.J.D. MacGregor\inst{2} \and I. Martin\inst{1} \and J.C. McGeorge\inst{2} \and F. Moschini\inst{1} \and F.D. Pacati\inst{4}\and M. Schwamb \inst{5} \and D. Watts \inst{6}
}                     
%
%
\institute{Kepler Centre for Astro and Particle Physics, Physikalisches
Institut, Universit\"{a}t T\"{u}bingen, D-72076 T\"{u}bingen, Germany \and
Department of Physics and Astronomy, University of Glasgow, Glasgow G12 8QQ, Scotland \and
Theoretical Nuclear Physics Laboratory, RIKEN Nishina Center, Wako 351-0198, Japan \and 
Dipartimento di Fisica Nucleare e Teorica dell'Universit\`{a} degli Studi di Pavia and Istituto Nazionale di Fisica Nucleare, Sezione di Pavia, I-27100 Pavia, Italy \and 
Institut f\"ur Kernphysik, Johannes-Gutenberg Universit\"at Mainz, Johann-Joachim-Becher-Weg 45, D-55099 Mainz, Germany \and 
School of Physics, University of Edinburgh, Edinburgh, United Kingdom
} 
\date{Received: date / Revised version: date}
%
\abstract{
After recent improvements to the Pavia model of two-nucleon knockout from
$^{16}$O with electromagnetic probes the calculated cross sections are compared to experimental data from such reactions. Comparison with data from a measurement of the $^{16}$O(e,e$'$pn) reaction show much better agreement between experiment and theory than was previously observed. In a comparison with recent data from a measurement of the $^{16}$O($\gamma$,pn) reaction the model over-predicts the measured cross section at low missing momentum.
\PACS{
      {21.30.Fe}{Forces in hadronic systems and effective interactions}   \and
      {21.60.-n}{Nuclear structure models and methods}   \and
      {25.20.Lj}{Photoproduction reactions}   \and
      {25.30.Fj}{Inelastic electron scattering to continuum}
     } 
} 
\maketitle
\section{Introduction}
\label{intro}

A major quest of nuclear physics is to understand how the properties
of nuclei arise from the underlying nucleon-nucleon (NN) interaction. 
 A useful starting point is given by the independent particle models (IPM),
in which protons and neutrons move freely in a common mean field. If one
accounts for spin-orbit effects and the average effects of the tensor
interaction~\cite{Otsuka05}, this approach explains the shell ordering
of most stable and dripline isotopes. However, this picture cannot
describe other basic observations, such as the strong fragmentation
of nuclear spectra and the corresponding quenching observed for absolute
spectroscopic factors~\cite{Lapikas93}.
The failure of the IPM arises from the correlated behaviour between nucleons,
which, at short inter-nucleon separations, is characterised by a strong
repulsion and, at intermediate to long range separations, by an attractive
interaction dominated by complicated tensor and spin-orbit terms. 
Thus, to understand nuclear structure a careful study of this correlated
behaviour is vital \cite{IPM_srcs,DB04}.

A direct method to study NN-correlations is by the use of two-nucleon knockout
reactions with an electromagnetic probe \cite{Boffi_book}. Proton-proton
and proton-neutron knockout reactions can act to probe the short range and
tensor components of the NN-interaction, respectively. Real and
virtual photons provide different and
complementary information on the reaction process. Real photons are only 
sensitive to transverse components of the interaction while virtual photons 
are sensitive to both the transverse and longitudinal components.

Electromagnetically induced two-nucleon
knockout reactions are driven by several processes. The coupling of the
(real or virtual) photon to either nucleon of a correlated pair via one-body hadronic 
currents can lead to the ejection of both nucleons from the nucleus. 
Interaction of the photon with two-body hadronic currents such as 
meson exchange currents (MEC) or isobar currents (IC) also contributes to the 
cross section. In addition final state interactions (FSI) between the two 
ejected nucleons and the recoil nucleus need to be taken into account. 
The relative importance of these different processes depends on the reaction 
type and kinematics.

The $^{16}$O nucleus is of particular interest regarding the study of correlated behaviour. Various theoretical models exist which attempt to describe these reactions 
\cite{16O_pn_advances_giusti,16O_eepp_ryckebusch_04,16O_gpn_anguiano} and there 
have been numerous measurements of two-nucleon knockout reactions using both 
real and virtual photons \cite{16O_eepn_middleton,16O_eepp_mami,16O_eepp_NIKHEF_1,16O_eepp_NIKHEF_2,16O_gNN_douglas,16O_gpn_lund,16O_gpn_canada}. 
In \cite{16O_eepn_middleton} the results from the first measurement of the 
$^{16}$O(e,e$'$pn)$^{14}$N reaction were reported and compared with theory.
Theoretical predictions for this reaction have been obtained
in \cite{16O_eepN_barbieri} by combining the self-consistent Green's functions
method for correlations and the Pavia model for the reaction mechanism. These
calculations suggest a strong sensitivity  of the cross sections to tensor 
correlations. However, they were unable to reproduce the 
shape or the magnitude of the data \cite{16O_eepn_middleton}.
These discrepancies sparked further developments to improve the reaction
model with respect to the
treatment of FSI \cite{16O_eepp_FSI_schwamb,16O_eeNN_FSI_schwamb}, of the 
two-body currents \cite{16O_eepp_delta_schwamb}, and of the centre-of-mass (CM) 
effects in the electromagnetic current operator  
\cite{16O_pn_advances_giusti,16O_eepp_CM}. 
 This paper presents a new comparison between experimental data and recent
calculations ~\cite{16O_pn_advances_giusti}, and shows that CM effects resolve
the discrepancy found in \cite{16O_eepn_middleton}.
We also show calculated cross sections for the $^{16}$O($\gamma$,pn)$^{14}$N reaction and compare
these to a recent measurement of this reaction \cite{Fede_thesis}.

\section{Theoretical calculations}
\label{theory}
 
The cross section of a reaction induced by a real or virtual photon, with 
momentum \mbox{\boldmath $q$}, where two nucleons are ejected from a nucleus can be written in
terms of the transition matrix elements of the  nuclear current operator 
between initial and final nuclear states.  Bilinear products of these matrix 
elements give the components of the hadron tensor and therefore the cross 
section  \cite{Boffi_book}.
For an exclusive process, where the residual nucleus is left in a discrete
eigenstate of its Hamiltonian, and under the assumption of a direct knock-out 
mechanism, the transition matrix elements contain three main ingredients: the 
two-nucleon overlap function between the ground state of the target and the 
final state of the residual nucleus, the nuclear current, and the two-nucleon 
scattering wave function \cite{tnko_giusti}.

The two-nucleon overlap function (TOF) contains information on nuclear 
structure and correlations. In \cite{16O_pn_advances_giusti} different 
treatments of correlations are compared, and produce dramatic differences
both in the shape and in the magnitude of the proton-neutron emission cross
sections. In particular, a crucial role is played by tensor correlations.
In the most refined approach, the TOF is obtained from a self-consistent 
calculation of the two-hole Green's function. In this case, the coupling
of nucleons and collective excitations of the system is calculated
microscopically from realistic NN forces. This is done employing the
Faddeev random phase approximation (FRPA) method discussed
in~\cite{Barb_FRPA_01,Barb_FRPA_07}. The long-range part of tensor
correlations is also included explicitly.
The TOF has been calculated in \cite{16O_eepN_barbieri} by partitioning 
the Hilbert space. Long-range correlations are evaluated using FRPA and
the Bonn-C NN-potential \cite{Bonn_C,Bonn_C1} in an appropriate
harmonic oscillator basis. The effects of short-range correlations, due to
the central and tensor part at high momenta, lie outside this space. Thus
they were added by computing the appropriate defect functions.

The nuclear current is the sum of a one-body and a two-body contribution. 
The one-body current includes the longitudinal charge term and the transverse 
convective and spin currents. The two-body current  is derived from a non 
relativistic reduction of the lowest-order Feynman diagrams with one-pion 
exchange and includes terms corresponding to the $\pi$-seagull and pion-in-flight 
diagrams, and to the diagrams with intermediate $\Delta$-isobar configurations.
Details of the nuclear current components can be found in 
\cite{16O_eepp_delta_schwamb,gNN_giusti,delta_w}. In comparison with the
previous calculations of \cite{16O_eepn_middleton}, the treatment of the 
two-body current has been improved using a more realistic regularised 
approach, which is consistent with the description of elastic NN-scattering 
data \cite{16O_pn_advances_giusti}

The two-nucleon scattering wave function contains the interaction of each one 
of the two outgoing nucleons with the residual nucleus, described in the model 
by an optical potential, as well as the mutual interaction of the two ejected 
nucleons (NN-FSI). In a simpler approach (DW) only the contribution of
FSI due to the optical potential is included, and the scattering state is 
written as the product of two uncoupled single particle distorted wave functions, 
eigenfunctions of a complex phenomenological optical potential which contains 
a central, a Coulomb, and a spin-orbit term.
In the more complete approach (DW-NN) the contribution of NN-FSI is also
included within the perturbative approach reported in 
\cite{16O_eepp_FSI_schwamb,16O_eeNN_FSI_schwamb}.

In comparison with earlier studies, a more complete treatment of CM 
effects has been included in the model 
\cite{16O_pn_advances_giusti,16O_pp_advances_giusti}. In the CM frame the 
transition operator becomes a two-body operator even in the case of a one-body 
nuclear current. As a consequence, the one-body current can give a contribution 
to the cross section of two-particle emission independently of correlations. 
These effects were not properly taken into account in the previous 
calculations for proton-neutron knockout \cite{16O_eepn_middleton,16O_eepN_barbieri}.
Accounting for CM effects is not trivial since the lack of orthogonality
between bound and scattering states (which are obtained from an 
energy-dependent optical potential) may give rise to spurious contributions to 
the calculated cross section. This issue has recently been overcome in 
\cite{16O_pn_advances_giusti} enforcing orthogonality between single particle initial and 
final states by means of the Gram-Schmidt procedure. 
The results show that the CM effects depend on kinematics. For the particular
case of  super-parallel kinematics, which were used in the measurement of the
$^{16}$O(e,e$'$pn)$^{14}$N reaction \cite{16O_eepn_middleton}, they enhance 
the contribution to the cross section which arises from the one-body currents. 
This effect is dramatic at low missing momentum and fully accounts for the previously observed \cite{16O_eepn_middleton}
discrepancy with experiment. The comparison between the 
$^{16}$O(e,e$'$pn)$^{14}$N data and the new calculations is reported in 
Sec.~\ref{res_eepn}.

\section{Experimental set-up}
\label{experiment}

\subsection{The $^{16}$O(e,e$'$pn)$^{14}$N reaction}
\label{exp_eepn}

A first measurement of $^{16}$O(e,e$'$pn)$^{14}$N reaction \cite{16O_eepn_middleton} was made at the electron scattering facility (3-spectrometer facility \cite{A1_blomqvist}) at MAMI, Mainz \cite{MAMI_B_walcher,MAMI_B_herminghaus}. Data were taken with an incoming electron beam of energy 855~MeV at currents of 10-20 $\mu$A. The beam was incident upon a waterfall target \cite{Waterfall_target} of thickness 74 mg cm$^{-2}$. The data were collected at energy and momentum transfers of 215~MeV and 316~MeV/\emph{c} where the ejected proton was detected in the forward direction, parallel to \textbf{\emph{q}}, with the ejected neutron detected in the backward direction, anti-parallel to \textbf{\emph{q}}, in so called ``super-parallel'' kinematics. The ejected proton and scattered electron were detected with Spectrometers A and B \cite{A1_blomqvist} of the 3-spectrometer set-up while the ejected neutron was detected using the Glasgow-T\"ubingen time-of-flight detector system \cite{TOF}. Further details about the experimental set-up and analysis of the data can be found in ref. \cite{16O_eepn_middleton}. The experimental resolution of the set-up was sufficient to distinguish groups of states in the residual nucleus but not good enough to separate individual states.

\subsection{The $^{16}$O($\gamma$,pn)$^{14}$N reaction}
\label{exp_gpn}

The $^{16}$O($\gamma$,pn)$^{14}$N reaction was measured at the Glasgow photon tagging facility \cite{tagger_MAMI_B,tagger_FP_MAMI_B} at MAMI, Mainz \cite{MAMI_B_walcher,MAMI_B_herminghaus}. An electron beam of energy 855 MeV used at a current of 50~nA was incident upon a 4 $\mu$m Nickel radiator to produce tagged Bremsstrahlung photons in the energy range 100 to 800~MeV. The Glasgow-tagger has an energy resolution of $~2$~MeV. The tagged photons, collimated to a diameter of $~18$~mm, were incident upon a target of 1~mm thickness. The target cell was filled with deuterated water and consisted of an Aluminium frame with polythene foil windows of 30~$\mu$m thickness which was orientated at an angle of $30^{\circ}$ with respect to the photon beam.

The ejected protons were detected in an array of four hyper-pure Germanium
detectors (HPGe) of the Edinburgh Ge6-Array \cite{Li6_halo_edinburgh}, each of which covered a solid angle of 59 msr and had a proton energy acceptance of 18 - 250~MeV. Pairs of double sided silicon strip detectors \cite{Fede_thesis} positioned in front of the HPGe detectors were were used to determine the trajectory of the ejected protons and reconstruct the reaction vertex. The ejected neutrons were detected at forward angles using the Glasgow-T\"ubingen time-of-flight detectors \cite{TOF}. Five neutron detector stands were used which covered an in-plane polar angular range of $6 - 53 ^{\circ}$ and a total solid angle of 146~msr. A pulseheight threshold of 5~MeV$_{ee}$ was used in the neutron detectors which resulted in a neutron kinetic energy threshold of $\approx$10~MeV. Full details of the experimental set-up and analysis of the data can be found in \cite{Fede_thesis}. The experimental resolution of the set-up was not able to resolve individual excited states in the residual $^{14}$N nucleus, only groups of states.

\section{Results}
\label{results}

\subsection{The $^{16}$O(e,e$'$pn)$^{14}$N reaction}
\label{res_eepn}

\begin{figure}
\resizebox{1\columnwidth}{!}{
  \includegraphics{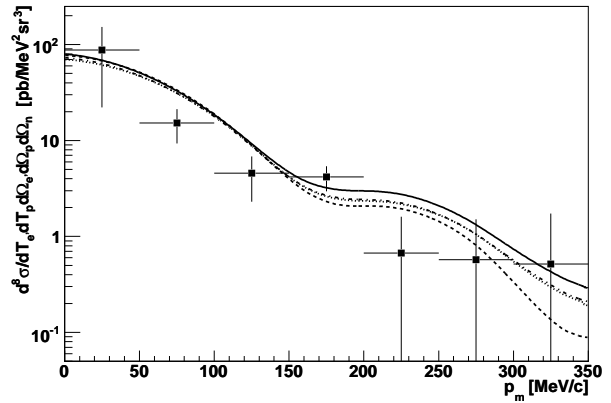}
}
\caption{The $^{16}$O(e,e$'$pn)$^{14}$N cross section shown as a function of the
missing momentum for events in the range 2 $\leq$ E$_{x}$ $\leq$ 9~MeV for energy and momentum transfers of 215~MeV and 316~MeV/\emph{c}. The
curves show the results from theoretical calculations of the cross section which includes
transitions to the first three excited states in $^{14}$N, 2.31~MeV (0$^{+}$),
3.95~MeV (1$^{+}$)  and 7.03 (2$^{+}$). The dashed line is calculated only with 
the one-body currents; the dotted line also includes the $\pi$-seagull term; 
the dashed dotted includes the one-body, $\pi$-seagull term and pion-in-flight 
terms and the solid line is for the complete cross-section including 
contributions from IC.}
\label{eepn_exp_comp_DW}
\end{figure}

Figure \ref{eepn_exp_comp_DW} shows the experimental and theoretical cross
sections of the $^{16}$O(e,e$'$pn)$^{14}$N reaction as a function of the 
absolute magnitude of the missing momentum
$\mbox{\boldmath $p$}_\mathrm{m}=\mbox{\boldmath
$q$}-\mbox{\boldmath$p'$}_\mathrm{p}-\mbox{\boldmath$p'$}_\mathrm{n}$,
where $\mbox{\boldmath$p'$}_\mathrm{p}$ and $\mbox{\boldmath$p'$}_\mathrm{n}$ 
are the momenta of the ejected nucleons.
The experimental cross section has been determined for a group of states  in the
residual $^{14}$N for an excitation energy range of 2 to 9~MeV. The theoretical
curves are the result of DW calculations and are the average cross section of 
calculations for the kinematic settings as given in \cite{16O_eepn_middleton}. 
The calculations represent 
the sum of contributions for transitions to three excited states in $^{14}$N: the 2.31\,MeV (0$^{+}$), 3.95~MeV (1$^{+}$) and 7.03~MeV (2$^{+}$) states. 

The theoretical curves of fig. \ref{eepn_exp_comp_DW} also show the
contributions of different terms of the nuclear current to the cross section. 
Cumulative contributions of the one-body, $\pi$-seagull, pion-in-flight and 
isobar currents are all shown. At low missing momentum the largest contribution
to the theoretical cross section is from one-body hadronic currents. 
Above $p_{\mathrm{m}}$ = 150~MeV/$c$ the $\pi$-seagull and ICs 
become increasingly more important with increasing $p_{\mathrm{m}}$. 
The pion-in-flight contribution is relatively small over the whole missing 
momentum range shown.

The shape of the experimental and theoretical cross sections in 
fig. \ref{eepn_exp_comp_DW} show reasonable agreement in that they both 
decrease roughly exponentially with increasing $p_{\mathrm{m}}$ and both show a 
flattening in the cross section at $p_{\mathrm{m}} \approx 175$\,MeV/$c$.
The magnitude of the two cross sections is in much better agreement compared 
to a previous comparison in \cite{16O_eepn_middleton} where the theoretical 
calculations under-predicted the experimental data at low $p_{\mathrm{m}}$. 
This improvement is due to the enhancement, at low $p_{\mathrm{m}}$,
of the contribution from the one-body currents produced by
the CM effects included in the present model \cite{16O_pn_advances_giusti}.

\begin{figure}
\resizebox{1\columnwidth}{!}{
  \includegraphics{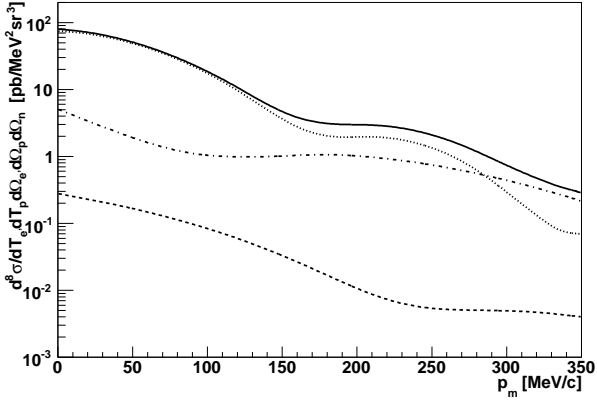}
}
\caption{Theoretical $^{16}$O(e,e$'$pn)$^{14}$N cross sections for energy and momentum transfers of 215~MeV and 316~MeV/\emph{c}. The
2.31~MeV (0$^{+}$), 3.95~MeV (1$^{+}$), 7.03 (2$^{+}$) and the three states
combined, represented by the dashed, dotted, dashed-dotted and solid lines
respectively. The plots are for the full cross section including the one-body, $\pi$-seagull, pion-in-flight and IC terms.}
\label{eepn_th_comp_3states}
\end{figure}

Figure \ref{eepn_th_comp_3states} shows a comparison of calculations of the 
full cross sections, including the one-body and two-body currents, for 
transitions to the three different excited states included in the curves of 
fig. \ref{eepn_exp_comp_DW}. The main strength in the cross section is predicted to come from transitions to the 3.95~MeV (1$^{+}$) state up to 
$p_{\mathrm{m}} \approx 290$\,MeV/$c$ where transitions to the 7.03 (2$^{+}$) 
state become dominant. The calculated contribution from transitions to the 2.31~MeV (0$^{+}$) state is at least an order of magnitude weaker, over the full $p_\mathrm{m}$ range shown, than those involving transitions to either of the other two states.

\begin{figure}
\resizebox{1\columnwidth}{!}{
  \includegraphics{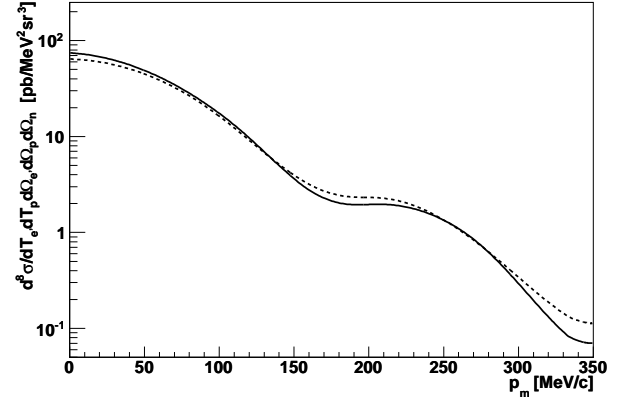}
}
\caption{Theoretical $^{16}$O(e,e$'$pn)$^{14}$N cross sections for transitions to the 3.95~MeV (1$^{+}$) excited state of $^{14}$N for energy and momentum transfers of 215~MeV and 316~MeV/\emph{c}. The solid curve uses the 
DW approach, the dashed line the DW-NN approach for FSI.}

\label{eepn_DW_DW-NN_comp}
\end{figure}

The calculations in figs. 1 and 2 are performed in the DW approach for FSI. NN-FSI
effects depend on kinematics and on the reaction type and are generally small
in proton-neutron emission  \cite{16O_eepp_FSI_schwamb,16O_eeNN_FSI_schwamb}.
For the $^{16}$O(e,e$'$pn)$^{14}$N reaction in the super-parallel kinematics 
NN-FSI are small but not negligible \cite{16O_pn_advances_giusti}. The effect of the mutual interaction between the two outgoing nucleons is shown in fig. \ref{eepn_DW_DW-NN_comp}, where the cross sections obtained in the DW and DW-NN approaches are compared for transitions to the 3.95~MeV (1$^{+}$) state in $^{14}$N. This one state dominates the reaction over nearly all of the 
measured $p_\mathrm{m}$ range. The effects of NN-FSI on the calculated cross section are 
relatively small. There is a slight decrease in cross section for 
$p_\mathrm{m} \leq 50$\,MeV/$c$ and a slight increase for 
$150 \leq p_\mathrm{m} \leq 225$\,MeV/$c$ and above $p_\mathrm{m} = 300$\,MeV/$c$. 
In general the calculations predict that NN-FSI have little importance
for the kinematics shown here. This fact justifies the perturbative treatment 
of NN-FSI.

\subsection{The $^{16}$O($\gamma$,pn)$^{14}$N reaction}
\label{res_gpn}

\begin{figure}
\resizebox{1\columnwidth}{!}{
  \includegraphics{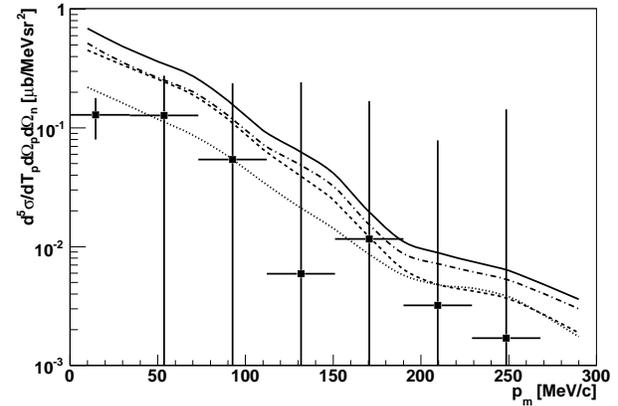}}
\caption{The $^{16}$O($\gamma$,pn)$^{14}$N cross section as a function of the
missing momentum for events in the range 2~$\leq$~E$_{x}$~$\leq$~10~MeV. The incident photon energy range was $150\leq E_{\gamma}\leq250$~MeV. The
curves show the theoretical cross section for
transitions to the 3.95~MeV (1$^{+}$) state. The dashed line is calculated with only one-body currents included; the dotted line also includes the $\pi$-seagull term; 
the dashed dotted includes the one-body, $\pi$-seagull term and pion-in-flight 
terms and the solid line is for the complete cross-section including 
contributions from IC.}
\label{gpn_DW_exp_comp}
\end{figure}

Figure \ref{gpn_DW_exp_comp} shows the cross section for the $^{16}$O($\gamma$,pn)$^{14}$N as a function of the absolute magnitude of the missing momentum, $p_\mathrm{m}$, of the reaction. The data are shown for an incident photon energy range of $150\leq E_{\gamma}\leq250$~MeV, proton in-plane azimuthal acceptance of $142 \leq \theta_{p} \leq 158^{\circ}$ and neutron in-plane azimuthal acceptance of $8 \leq \theta_{n} \leq 32^{\circ}$. The experimental cross section has been determined for a group of states in the recoiling $^{14}$N nucleus for an excitation energy range of 2 to 10~MeV. Figure \ref{gpn_DW_exp_comp} also shows the results of DW theoretical calculations for the reaction. The curves are for transitions to the 3.95~MeV (1$^{+}$) state which is believed to dominate the cross section as in \cite{16O_gpn_lund} and have been averaged over the kinematic settings which cover the acceptance of the experimental data.

The theoretical curves of fig. \ref{gpn_DW_exp_comp}  show the contributions of different terms of the nuclear current to the cross section. Cumulative contributions of the one-body, $\pi$-seagull, pion-in-flight and 
isobar currents are all shown, see the caption of fig. \ref{gpn_DW_exp_comp} for details. At
low $p_\mathrm{m}$ the largest contribution to the theoretical cross section is
from one-body hadronic currents. The inclusion of the $\pi$-seagull term causes
a decrease in calculated cross section until roughly $p_\mathrm{m} = 200$~MeV/$c$
where it has very little effect. The further inclusion of the pion-in-flight
contributions increases the cross section to roughly the same strength of the
one-body hadronic current cross section for $p_\mathrm{m} < 100$~MeV/$c$ after
which point it increases the calculated cross section
relative to the one-body hadronic currents alone. The inclusion of ICs increases the calculated cross section for the whole $p_\mathrm{m}$ range shown.

Both the theoretical and experimental cross sections shown in fig. 4 show a similar trend of
falling roughly exponentially with increasing $p_\mathrm{m}$. The theory over-predicts the experimental data at low $p_\mathrm{m}$. Fig. \ref{gpn_DW_exp_comp} suggests that the discrepancy may decrease with increasing $p_\mathrm{m}$ but more accurate measurements are necessary to confirm this.

\begin{figure}
\resizebox{1\columnwidth}{!}{
  \includegraphics{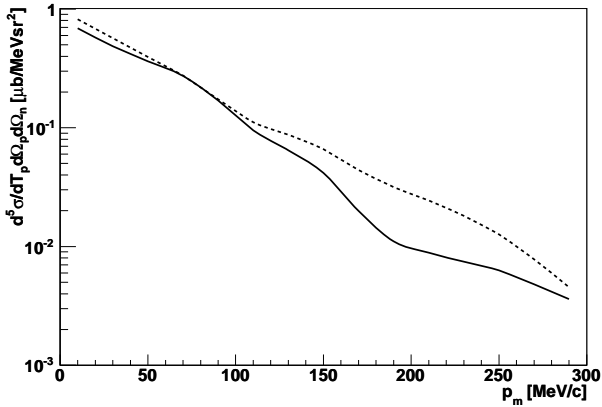}}
\caption{Calculations of the $^{16}$O($\gamma$pn)$^{14}$N cross section for transitions to the 3.95~MeV (1$^{+}$) excited state of $^{14}$N. The solid curve uses the 
DW approach, the dashed line the DW-NN approach for FSI.}

\label{gpn_DW_DW-NN_comp}
\end{figure}

The effect of the mutual interaction between the two outgoing nucleons for the ($\gamma$,pn) reaction is shown in fig.~\ref{gpn_DW_DW-NN_comp}. Theoretical cross sections were obtained using the DW and DW-NN approaches for transitions to the 3.95~MeV (1$^{+}$) state in $^{14}$N. At low $p_\mathrm{m}$ the effects of NN-FSI on the calculated cross section are very small. From about $p_\mathrm{m} = 100$~MeV/$c$ the importance of NN-FSI increases until roughly $p_\mathrm{m} = 200$~MeV/$c$ after which their importance again diminishes. This is in contrast to what was seen for the (e,e$'$pn) reaction where NN-FSI had very little effect on the calculated cross section. The inclusion of NN-FSI increases the theoretical cross
section at high $p_\mathrm{m}$ which, however, remains well within the statistical error bars associated with the data points in this region.

\section{Conclusions}
\label{conclusions}

An improved treatment of centre-of-mass effects in the electromagnetic current
operator of the Pavia model has resulted in closer agreement with experimental
data for the $^{16}$O(e,e$'$pn)$^{14}$N reaction with both the shape and
magnitude of the experimental cross section being well described.
 
A further comparison of the improved model
with data from a recent measurement of the $^{16}$O($\gamma$,pn)$^{14}$N reaction
showed a similar shape with $p_\mathrm{m}$ but over-predicted the strength of the
measured cross section at low $p_\mathrm{m}$. However, more accurate data would be
required in order to draw a more detailed interpretation from this
comparison.

\section*{Acknowledgments}
The authors would like to thank the staff of the Institut f\"{u}r Kernphysik
in Mainz for providing the facilities in which these experiments took
place. This work was sponsored by the Deutsche Forschungsgemeinschaft and the UK Science and Technology Facilities Council (STFC).

\bibliographystyle{unsrt}

\end{document}